\documentclass[twocolumn,showpacs]{revtex4}
\usepackage{graphicx}
\graphicspath{{pict/}{}}

\usepackage{bm}%

\newcounter{Fig}

\newcommand{\be}{\begin{equation}}
\newcommand{\ee}{\end{equation}}

\begin{document}

\title{Interface solitons in quadratically nonlinear photonic lattices}

\author{Zhiyong Xu$^1$, Mario I. Molina$^2$, and Yuri S. Kivshar$^1$}

\affiliation{$^1$Nonlinear Physics Center, Research School of Physics
and Engineering, Australian National University, Canberra ACT 0200,
Australia\\
$^2$Departmento de F\'{\i}sica, Facultad de Ciencias,
Universidad de Chile, Casilla 653, Santiago, Chile}

%\date{\today}

\begin{abstract}
We study the properties of two-color nonlinear localized modes which may exist at the interfaces separating two different periodic photonic lattices in quadratic media, focussing on the impact of phase mismatch of the photonic lattices on the properties, stability, and threshold power requirements for the generation of interface localized modes. We employ both an effective discrete model and continuum model with periodic potential and find good qualitative agreement between both models. Dynamics excitation of interface modes shows that, a two-color interface twisted mode splits into two beams with different escaping angles and carrying different energies when entering a uniform medium from the quadratic photonic lattice. The output position and energy contents of each two-color interface solitons can be controlled by judicious tuning of the lattice parameters.
\end{abstract}


\maketitle

\section{Introduction}

Electromagnetic surface waves are waves localized at an
interface separating either \emph{two homogeneous} (one of them
has to be surface-active, i.e., exhibiting a negative
permittivity~\cite{Borstel}) or \emph{homogeneous} and
\emph{periodic} dielectric media~\cite{Yeh_APL_78}.  In addition,
nonlinear dielectric media can support nonlinear guided waves
localized at or near the surfaces, and different types of
nonlinear guided waves in planar waveguides have been studied
extensively about 20 years ago~\cite{review,nw1}. Recently, the interest in the study
of electromagnetic surface waves has been renewed after the
theoretical prediction~\cite{OL_george} and subsequent
experimental demonstration~\cite{PRL_george} of
nonlinearity-induced self-trapping of light near the edge of a
one-dimensional waveguide array with self-focusing nonlinearity,
that can lead to the formation of a {\it discrete surface soliton}. A
related effect of light localization and the formation of surface
gap solitons have been predicted theoretically and observed
experimentally for defocusing periodic nonlinear
media~\cite{PRL_kartashov,PRL_canberra}. In addition, the concept
of nonlinear surface and gap solitons has been extended to the
case of an interface separating two different nonlinear periodic
media~\cite{OL_inter,OE_poly,PLA_mario,OE_stegeman01}.

Surface solitons are usually considered for one-frequency modes propagating in cubic or saturable nonlinear media.
However, multicolor discrete solitons in quadratically nonlinear lattices have been studied
theoretically in both one- and two-dimensional lattices~\cite{chi2_p1,chi2_p2,chi2_p3,chi2_p4}
irrespective to the surface localization effects. Only Siviloglou et al.~\cite{OE_stegeman} studied
discrete quadratic surface solitons experimentally in periodically poled lithium
niobate waveguide arrays, and they employed a discrete model with decoupled waveguides
at the second harmonics to model some of the effects observed experimentally.

More elaborated theory of one-dimensional surface solitons in truncated quadratically nonlinear photonic lattices, the so-called {\it two-color surface lattice solitons}, has been developed recently by Xu and Kivshar~\cite{chi2_our} who analyzed the impact of the phase mismatch on the existence
and stability of nonlinear parametrically coupled surface modes, and also found novel classes of
one-dimensional two-color twisted surface solitons which are stable in a large domain of their existence.

The purpose of this paper is twofold. First, we extend the analysis of two-color surface solitons to the case
of two semi-infinite photonic lattices with quadratic nonlinearities. We study, for the first time
to our knowledge, two-color interface solitons in photonic lattices with quadratic nonlinear response.
We analyze the effect of mismatch on the existence, stability, and generation of such novel surface states.
 Second, for the analysis outlined above we employ two different approaches widely used in the literature: The coupled-mode theory described by the discrete parametrically coupled equations for the fundamental and second-harmonic fields, and also
 the continuous model with a periodic potential, and demonstrate that both models give the same qualitative results.

The paper is organized as follows. In Sec.~II we discuss the two-color interface localized modes in the framework of a discrete
model where the interface is modeled by a jump of the propagation constant, assuming that the matching conditions are nearly
satisfied for both semi-infinite lattices. Section~III is devoted to the analysis of a more general case described by a
continuous model with a periodically varying potential. Thus we consider the light beam propagating along the interface between
two dissimilar optical lattices imprinted in quadratic nonlinear media. Besides the study of the properties of interface solitons,
we demonstrate the manipulation of surface solitons by tuning the lattice and waveguide parameters. Finally, Sec.~IV concludes the
paper.

\section{Discrete model}

We consider the propagation of light in a one-dimensional photonic
lattice of a finite extent imprinted in quadratic nonlinear media,
which involves the interaction between fundamental frequency (FF)
and second-harmonic (SH) waves. Light propagation is described by
the following coupled nonlinear discrete
equations~\cite{chi2_p1,chi2_our}
\begin{eqnarray}
\label{eq:model}
&& i\frac{d u_{n}}{d
z}+C_u (u_{n+1} + u_{n-1}) +2\gamma u_{n}^{*}v_{n} \exp(+i\beta z)=0, \nonumber \\
&& i\frac{d v_{n}}{d z} + C_v (v_{n+1}+ v_{n-1})
+\gamma u_{n}^2 \exp(-i\beta z)=0,\label{eq:1}
\end{eqnarray}
where $u_{n}$ and $v_{n}$ are the normalized amplitudes of the FF and SH waves, respectively,
$C_u$ and $C_v$ are the coupling coefficients, $\gamma$ characterizes the second-order nonlinearity,
and $\beta$ is the effective mismatch between the two harmonics.

\begin{figure}[h]
\centering
 \includegraphics[width=4.2cm]{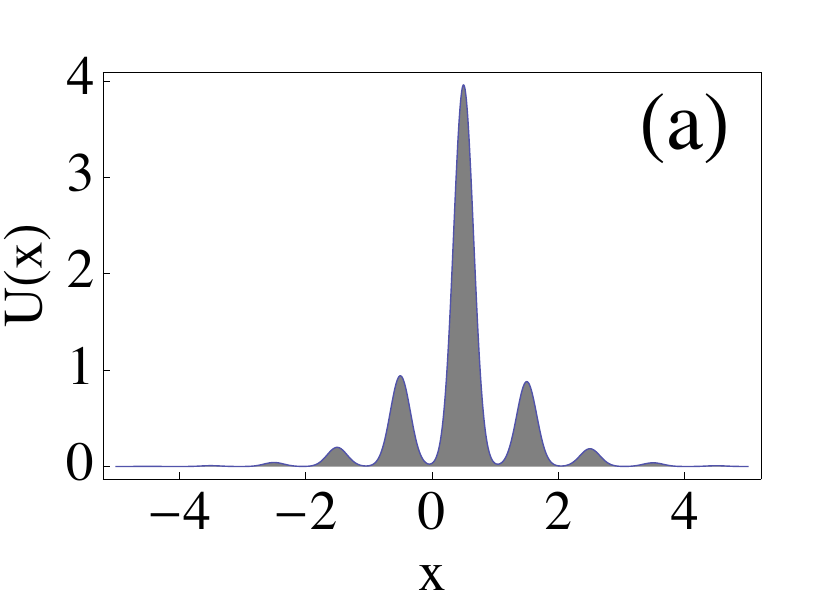}
  \includegraphics[width=4.2cm]{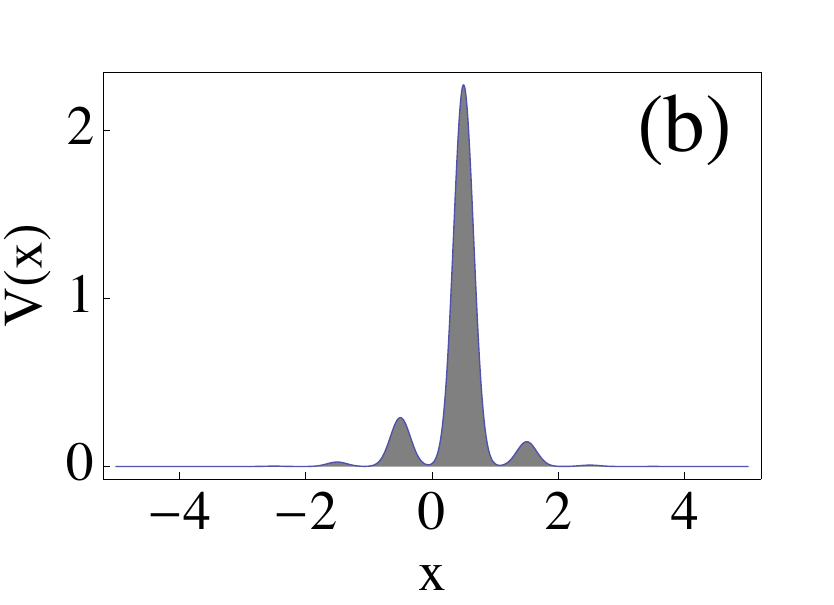}\\
   \includegraphics[width=4.2cm]{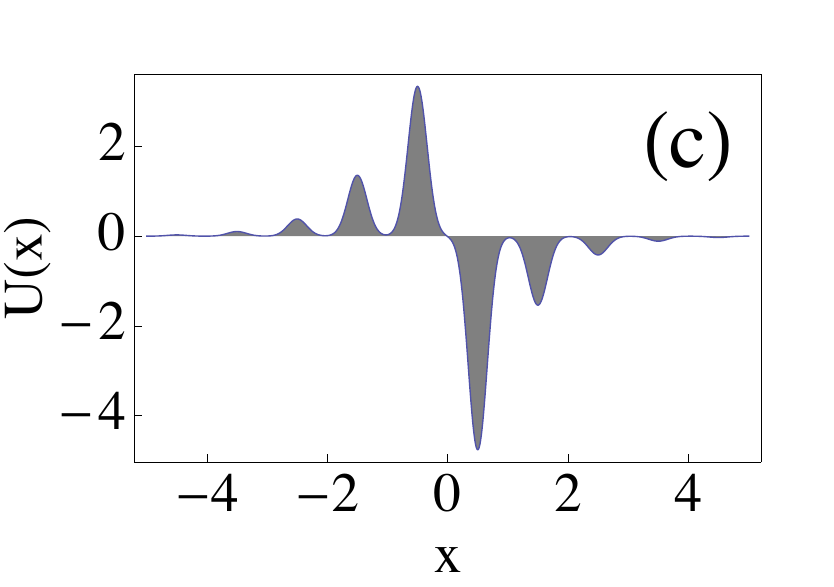}
    \includegraphics[width=4.2cm]{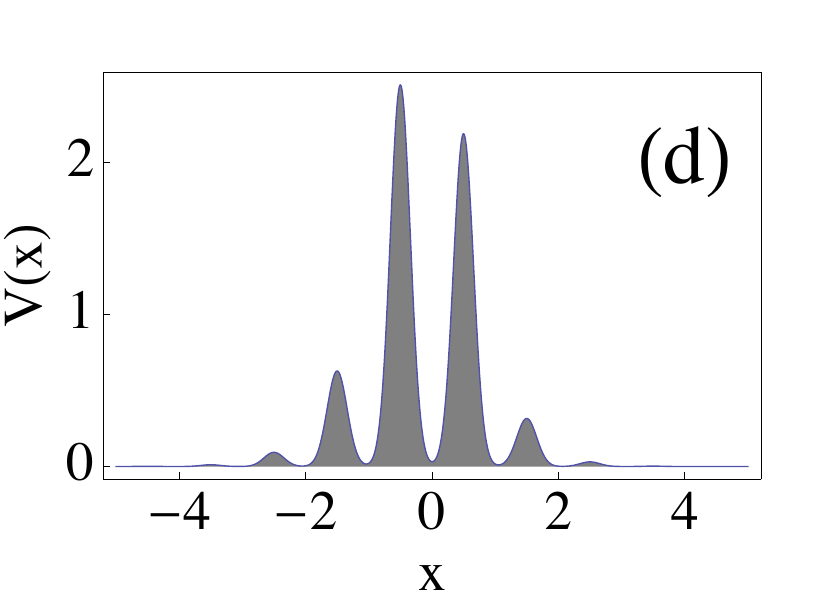}
\caption{Examples of two-color interface solitons for
$C_u=1$, $C_v=0.5$ and (a,b) $b=5$, $\beta_1=3$, $\beta_2=-3$, and
(c,d) $b=4$, $\beta_1=3$, $\beta_2=-3$. Shown are the amplitudes
of the fundamental (a,c) and second-harmonic (b,d) components.
Case (a,b) corresponds to a fundamental interface mode, while (c,d)
corresponds to an interface ``twisted'' mode.}
\label{fig1}
\end{figure}

We look for stationary two-mode solutions of Eq.~(\ref{eq:1})
in the form, $u_{n}(z) = U_{n} \exp(ibz)$ and $v_{n}(z) = V_{n}
\exp(2ibz - i \beta z)$, where $b$ is the propagation constant, and obtain the nonlinear algebraic
equations for the (real) mode amplitudes, $U_{n}$ and $V_{n}$,
\begin{eqnarray}
-b\ U_{n} + C_{u}(U_{n+1}+ U_{n-1}) +2 \gamma U_{n} V_{n} & =& 0,\nonumber\\
 -2b\ V_{n} + C_{v}(V_{n+1}+ V_{n-1}) + \beta_n V_{n}+\gamma U_{n}^{2} &= &0,
 \label{eq:2}
\end{eqnarray}
where $\beta_{n}=\beta_{1}$ on the left side of the interface, and $\beta_{n}=\beta_{2}$ on the right side. We solve Eqs.~(\ref{eq:2}) numerically, via the well-known
multi-dimensional Newton-Raphson method, using the results of the
anti-continuum limit as initial conditions for the algorithm: When
$b \gg {\rm Max}[\beta_1; \beta_2]$, it is easy to obtain, $V_n
\approx b/(2\gamma)$ and $U_n \approx b/\gamma$.
Figures 1(a,b) show an example of the simplest interface mode,
where the fields propagating along  the waveguides are presented as a superposition of the waveguide modes, $U(x)=\sum_{n} U_{n} \phi^{(ff)}(x-n)$,
where $\phi^{(ff)}(x)$ is the single FF mode centered on guide $n$; similarly for $V(x)$. Using the same numerical approach and starting from the anti-continuum limit, we find several other families of
two-color localized modes located at the interface. These modes provide a  generalization of the surface modes known
for truncated one-dimensional lattices located at different distances from the edge, and corresponding to a crossover
between the interface and bulk discrete solitons as discussed earlier~\cite{OL_molina}. In addition, we find novel
classes of the so-called interface twisted modes, and an example of such mode  is shown in Figs.~\ref{fig1}(c,d). In general the spatial profile of these modes are asymmetrical for $\beta_{1}\neq\beta_{2}$.

For given values of $C_u$, $C_v$, $\beta_1$, and $\beta_2$, we
compute the minimum value of the propagation constant $b$ and the
power
\begin{equation}
P=P_u +P_v=\sum_{n}(|u_{n}|^2 + 2|v_{n}|^2),
\label{power}
\end{equation}
needed to create a localized mode at one of the interface sites.
Figures 2(a,b) show examples of $P_{\rm min}$ curves for two
cases. In the first case, when no interface is present, there is
no minimum power to generate a localized mode for small values of
$\beta$. Outside this interval, power increases more or less
symmetrically with increase in $|\beta|$. On the other hand, in
the presence of an interface we observe that the generation of an
interface localized mode require a finite power, as shown in
Fig.~2(b). The numerically-observed stability of these modes seems
to be in agreement with the stability results of the continuous
model (see section III).

\begin{figure}[h]
\centering
 \includegraphics[width=4.2cm]{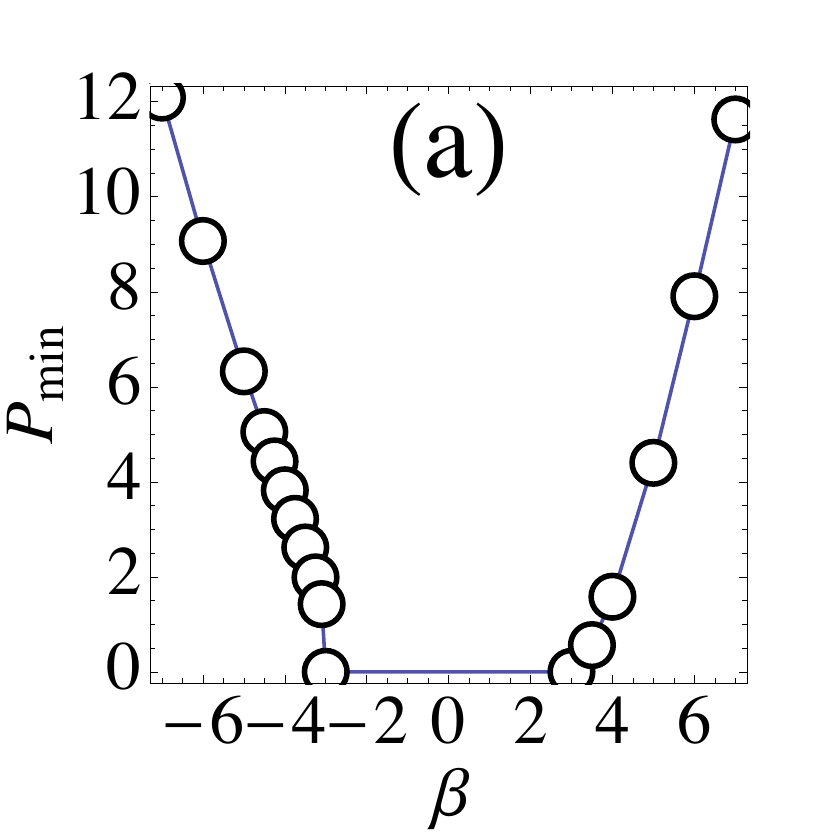}
  \includegraphics[width=4.2cm]{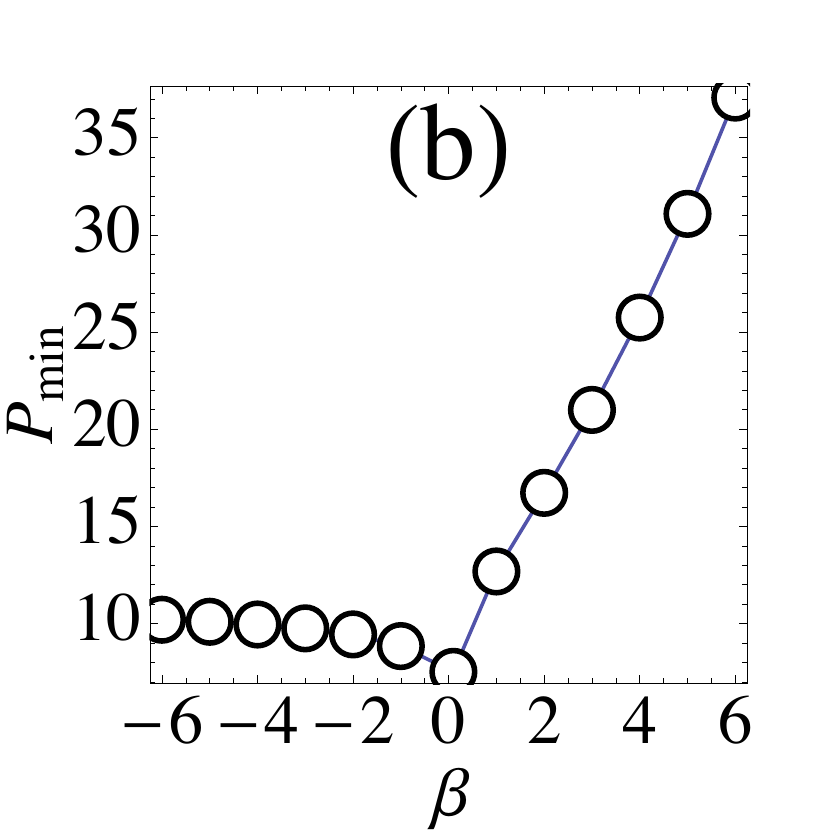}
\caption{Minimum power $P_{\rm min}$ of the simplest two-color interface soliton
for $C_u=1$, $C_v=0.5$, and (a) $\beta_1=\beta_2=\beta$, (b)
$\beta_1=0$ and $\beta_2=\beta$.}
\label{fig2}
\end{figure}

\begin{figure}[h]
\centering
 \includegraphics[width=4.2cm]{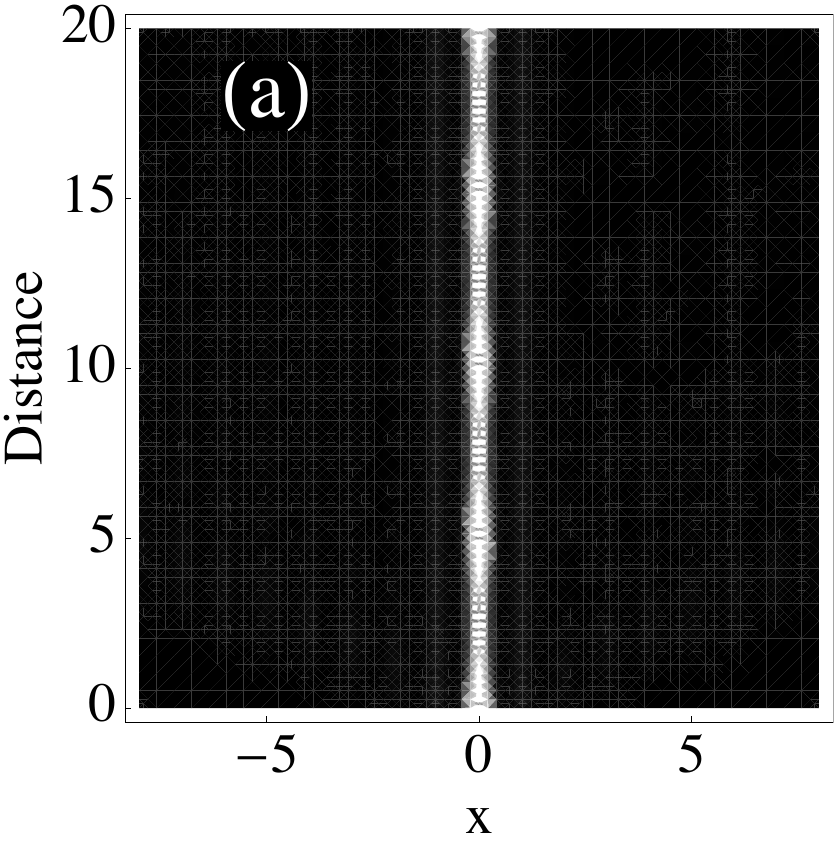}
  \includegraphics[width=4.2cm]{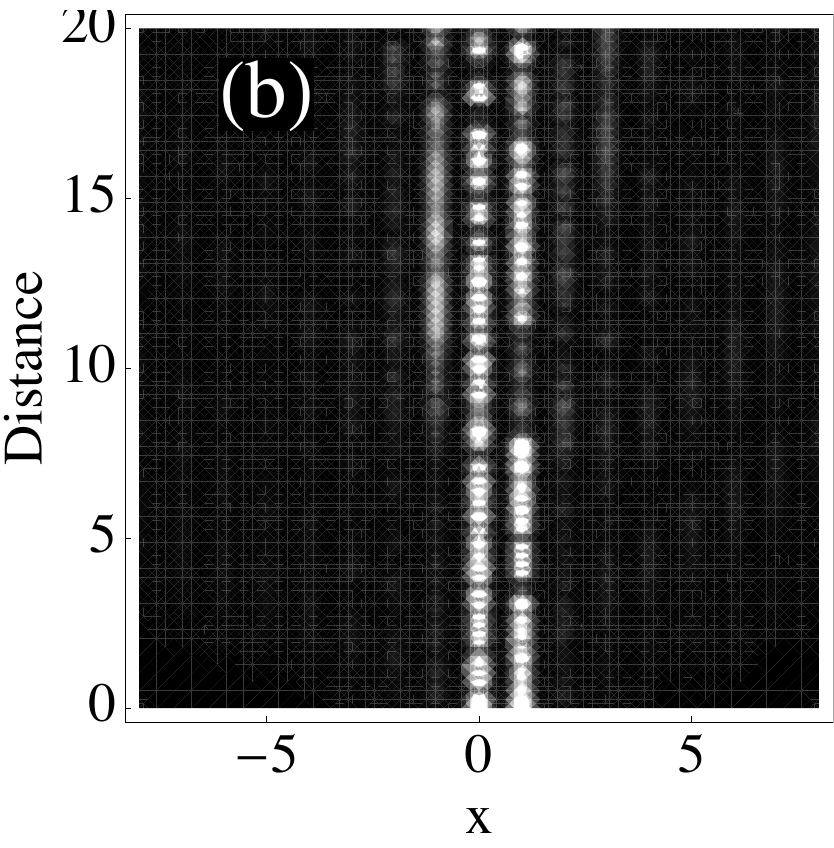}
\caption{Dynamical excitation of interface modes. Power content in the FF field for (a) Simple odd mode and (b) Twisted mode. Only FF field is excited initially. Here $C_{u}=1, C_{v}=0.5, \beta_{1}=3, \beta_{2}=-3$.}
\label{newfig3}
\end{figure}

The dynamical excitation of the interface mode can be achieved by launching all of the initial power in one of the guides adjacent to the interface. When both,  FF and SH fields are initially present, a localized mode is formed if the power is strong enough. In the more experimentally relevant case when only the FF field is initially present, the threshold power increases significantly. After  the interface mode has been established, the power exchange between the FF and SH fields proceeds along the longitudinal direction with a spatial period that decreases with increasing initial power, for $\beta_{1}, \beta_{2}$ fixed. On the other hand, for a fixed value of initial power in the FF field, enough to excite the interface mode, the spatial period for power exchange between the fields decreases with an increase in the value of $|\beta|$ at the initial launching guide. Figure \ref{newfig3} shows an example of the dynamical excitation of a simple odd mode and a twisted mode, where only the FF field is excited initially. Internal oscillations of the field as well as asymmetrical power sharing for the twisted mode are apparent.

\section{Continuous model with a periodic potential}

\begin{figure}[htb]
\centerline{\includegraphics[scale=.80, angle=270]{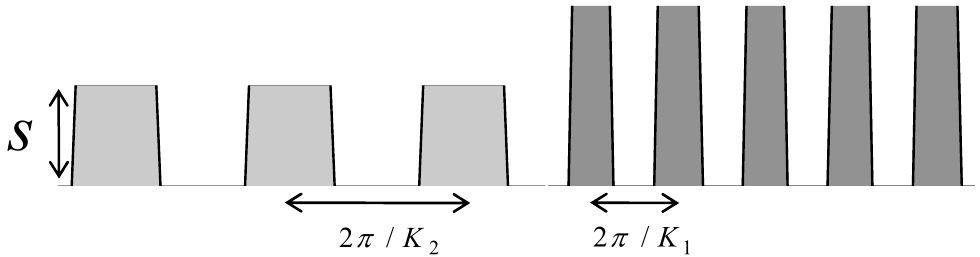}}
\caption{Interface structure created by two different quadratic lattices with tunable guiding parameters.}
\label{lattice}
\end{figure}

In this section, we extend the analysis to a more general
model (a continuous model with a periodically varying potential)
to describe propagation of light at the interface between two
dissimilar optical lattices imprinted in quadratic nonlinear
media, which involves the interaction between fundamental
frequency and second-harmonic waves. Light propagation is
described by the following coupled nonlinear equations
\cite{chi2_our}
\begin{eqnarray}\label{eq:model}
&& i\frac{\partial u}{\partial
z}=\frac{d_{1}}{2}\frac{\partial^{2}u}{\partial x^{2}}-
u^{*} v \texttt{exp}(-i\beta z)-pR(x)u, \nonumber \\
&&i\frac{\partial v}{\partial
z}=\frac{d_{2}}{2}\frac{\partial^{2}v}{\partial x^{2}}-
u^{2}\texttt{exp}(i\beta z)-2pR(x)v,
\end{eqnarray}
where $u$ and $v$ represent the normalized complex amplitudes of
the FF and SH fields, $x$ and $z$ stand for the normalized
transverse and longitudinal coordinates, respectively, $\beta$ is
the phase mismatch, and $d_{1}=-1$, $d_{2}=-0.5$; $p$ is the
lattice depth; the function $R(x)=s[1-\texttt{cos}(K_{2} x)]$ at
$x<0$ and $R(x)=1-\texttt{cos}(K_{1} x)$ at $x\geq 0$ describes
the profile of a waveguide array formed with two dissimilar lattices with
modulation frequencies $K_{1}$ and $K_{2}$, respectively, where
$s$ is the relative lattice depth for the left-side one [a typical
profile for such tunable waveguide array is shown in
Fig.~\ref{lattice}]. In typical quadratic nonlinear crystals, for
a beam width of $\sim 15 \mu \texttt{m}$, the distance $z$ in the
range $0-30$ corresponds to a few centimeters, and the peak
intensities will be in the range of
$0.1-10\texttt{GW}/\texttt{cm}^{2}$ for the formation of lattice
solitons at wavelengths $\lambda=1 \mu \texttt{m}$; a refractive
index modulation depth of the order of $10^{-4}$ corresponds to
the lattice depth $p\sim1$ \cite{chi2_our}. The system of
Eq.~(\ref{eq:model}) admits several conserved quantities including
the power $P=\int_{-\infty}^{\infty}(|u|^{2}+|v|^{2})dx$.

The stationary solutions for the lattice-supported interface
solitons can be found in the form
$u(x,z)=w_{1}(x)\texttt{exp}(ib_{1}z)$ and
$v(x,z)=w_{2}(x)\texttt{exp}(ib_{2}z)$, where $w_{1,2}(x)$ are
real functions, and $b_{1,2}$ are real propagation constants
satisfying $b_{2}=\beta+2 b_{1}$. Families of interface solitons are
determined by the propagation constant $b_{1}$, the lattice depth
$p$ and $s$, and the phase mismatch $\beta$. For simplicity, we
set the modulation parameter $K_{1}=4$ and vary $K_{2}$. To
analyze stability we examine perturbed solutions
$u(x,z)=[w_{1}(x)+U_{1}(x,z)+iV_{1}(x,z)]\texttt{exp}(ib_{1}z)$
and
$v(x,z)=[w_{2}(x)+U_{2}(x,z)+iV_{2}(x,z)]\texttt{exp}(ib_{2}z)$,
where real parts $U_{1,2}$ and imaginary parts $V_{1,2}$ of
perturbation can grow with complex rate $\delta$. The
linearization of Eq.~(\ref{eq:model}) around $w_{1,2}$ yields the
following eigenvalue problem:
\begin{eqnarray}\label{eq:model2}
&&\delta U_{1}=\frac{d_{1}}{2}\frac{\partial^{2}V_{1}}{\partial x^{2}}-
(w_{1}V_{2}-w_{2}V_{1})-pRV_{1}+b_{1}V_{1}, \nonumber \\
&& \delta V_{1}=-\frac{d_{1}}{2}\frac{\partial^{2}U_{1}}{\partial x^{2}}+
(w_{1}U_{2}+w_{2}U_{1})+pRU_{1}-b_{1}U_{1}, \nonumber \\
&& \delta U_{2}=\frac{d_{2}}{2}\frac{\partial^{2}V_{2}}{\partial x^{2}}-
2w_{1} V_{1}-2pRV_{2}+b_{2}V_{2}, \nonumber \\
&& \delta V_{2}=-\frac{d_{2}}{2}\frac{\partial^{2}U_{2}}{\partial
x^{2}}+ 2w_{1}U_{1}+2pRU_{2}-b_{2}U_{2},
\end{eqnarray} which we solve numerically to find the growth rate $\delta$.

\begin{figure}[htb]
\centerline{\includegraphics[width=7.5cm]{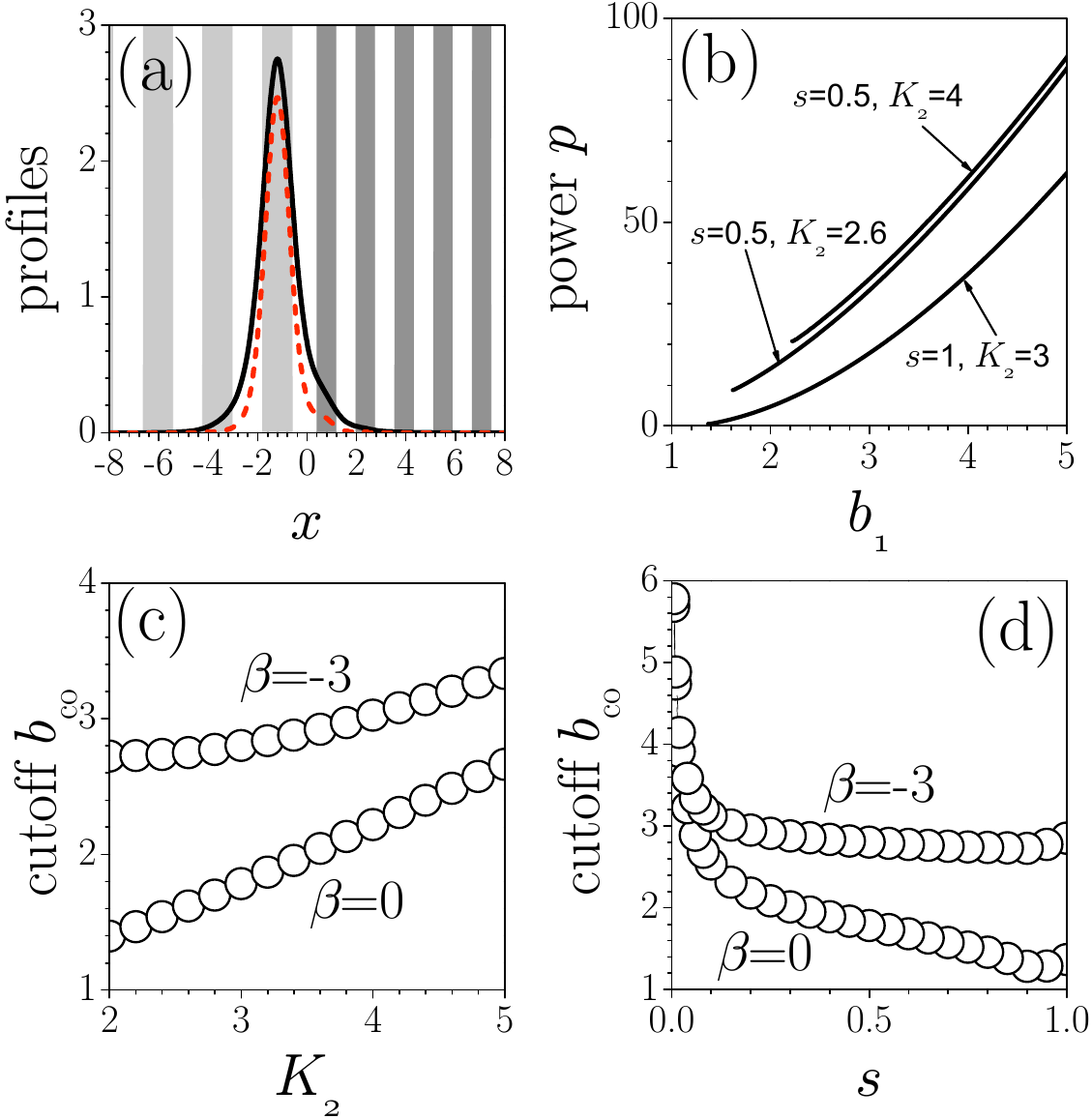}}
\caption{(Color online) Two-color odd interface solitons (a) FF (black solid curve) and SH (red dashed curve) field distributions of two-color odd interface solitons with $b_{1}=4$ at $s=0.5$ and $K_{2}=2.6$. (b) Power versus propagation constant for different parameters as marked in the plot, In (a,b) phase matching $\beta=0$. (c, d) Cutoffs of the propagation constant $b_{1}$ versus (c) the lattice period $K_{2}$ at $s=0.5$ and (d) the relative lattice depth $s$ at $K_{2}=3$ for different $\beta$. }
\label{fig2}
\end{figure}

The tunable waveguide array shown in Fig.~\ref{lattice} supports
rich families of two-color interface lattice solitons. In
semi-infinite waveguide arrays ($s=0$) the properties of
two-color surface solitons have been
investigated~\cite{chi2_our}. Here we are interested in the case
$s\neq0$ thus we can show how to manipulate two-color lattice
interface solitons by tuning the lattice and waveguide parameters
$s$ and modulation frequency $K_{2}$. The simplest example of a
two-color interface mode (odd soliton) is shown in
Fig.~\ref{fig2}(a), from which one notices that the peak of this
mode coincides with one of the maxima of the lattices. As shown in
Fig.~\ref{fig2}(b) its total power is almost a monotonic function
of the propagation constant. It it important to note that the
existence domain of two-color interface solitons is determined by
the waveguide parameters, thus the cutoff value of propagation
constant varies depending on the lattice period
[Fig.~\ref{fig2}(c)] and depth [Fig.~\ref{fig2}(d)]. A linear
stability analysis reveals that these odd solitons are stable
almost in the entire domain of their existence.

\begin{figure}[htb]
\centerline{\includegraphics[width=7.5cm]{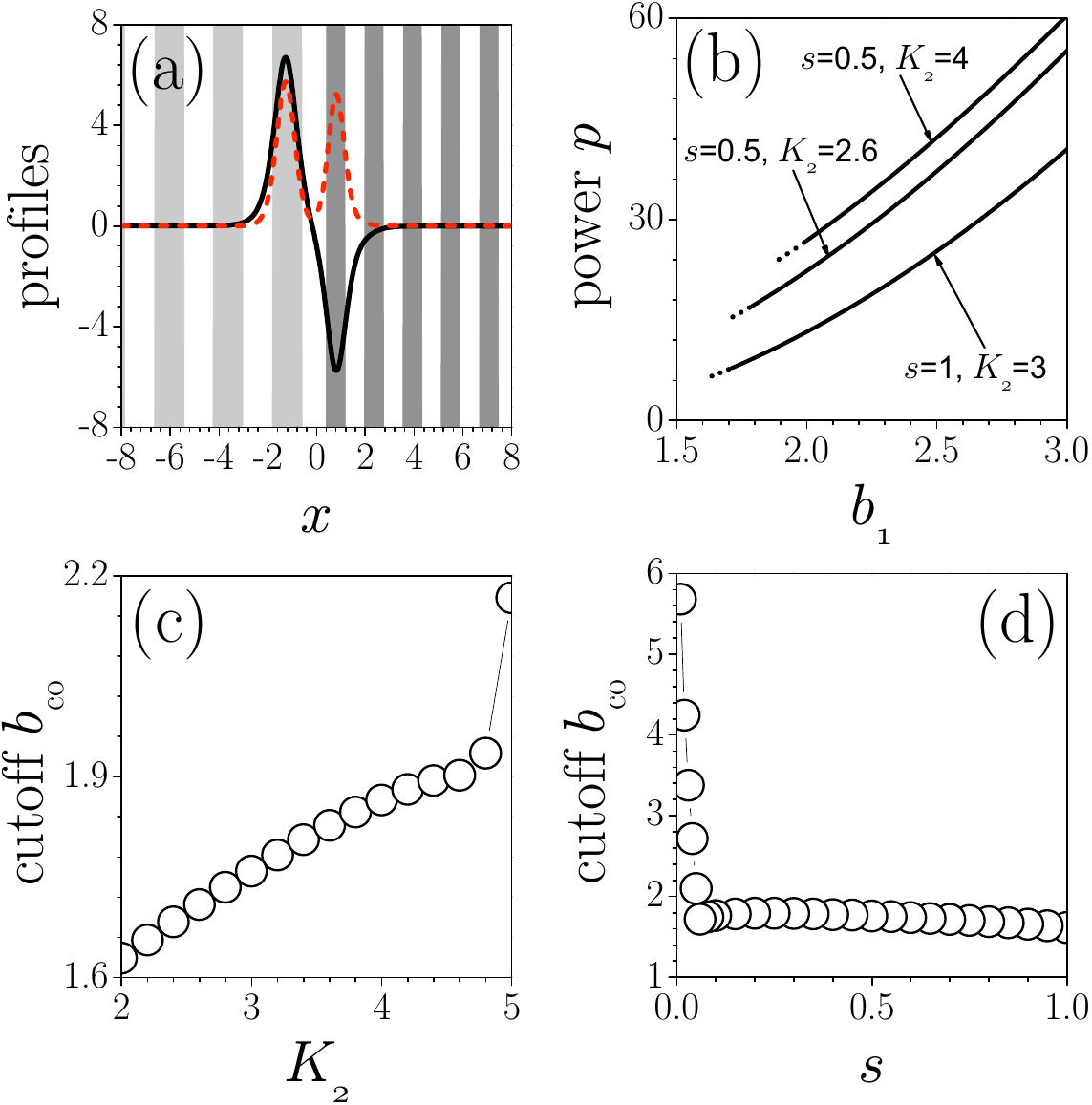}}
\caption{(Color online) Two-color twisted interface solitons (a) FF (black solid curve) and SH (red dashed curve) field distributions with $b_{1}=4$ at $s=0.5$ and $K_{2}=2.6$. (b) Power versus propagation constant for different parameters as marked in the plot [Solid and dotted curves correspond to stable and unstable branches, respectively]. (c, d) Cutoffs of the propagation constant $b_{1}$ versus (c) the lattice period $K_{2}$ at $s=0.5$ and (d) the relative lattice depth $s$ at $K_{2}=3$. Here phase matching $\beta=0$.}
\label{fig3}
\end{figure}

\begin{figure}[htb]
\centerline{\includegraphics[width=8.44cm]{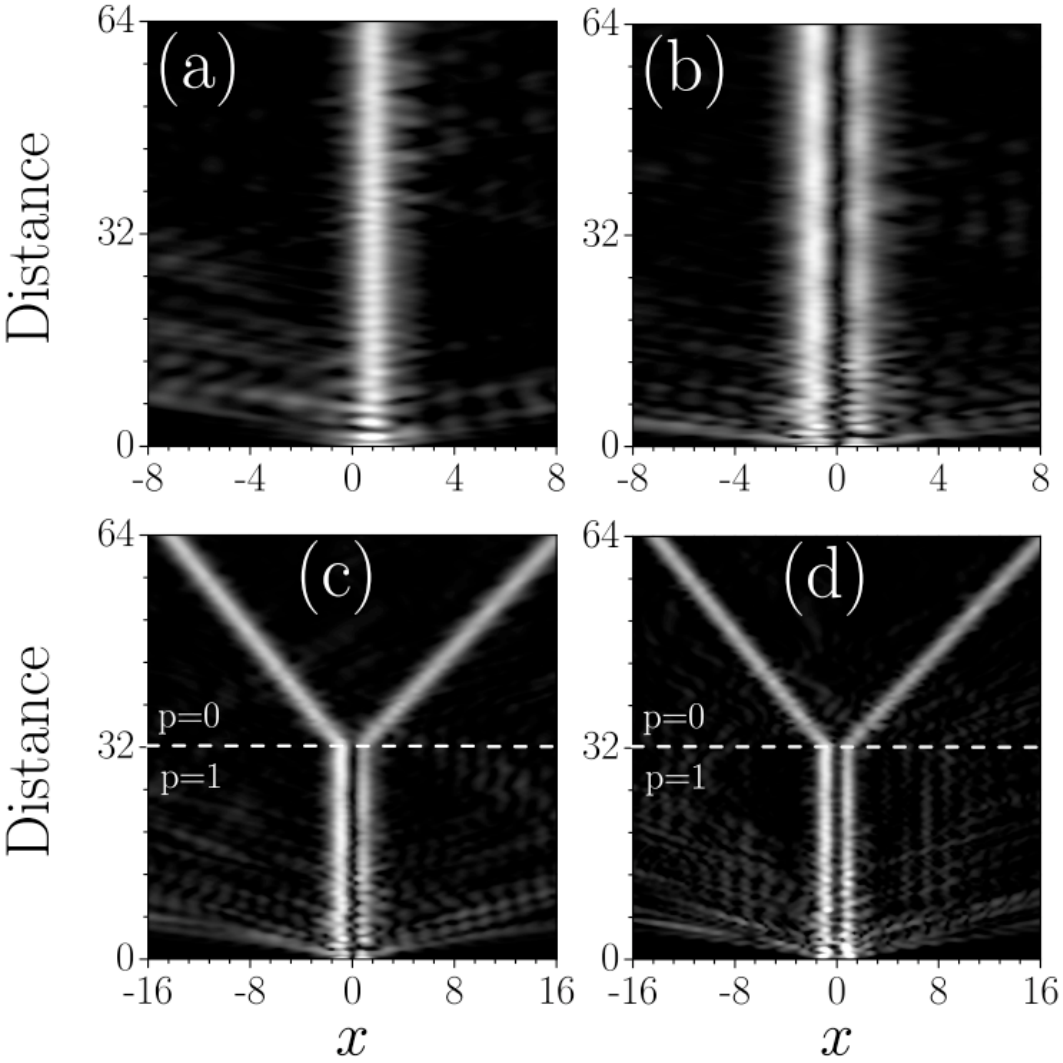}}
\caption{Excitation of interface solitons from only FF field as input beam: FF field distributions of generated odd
(a) and twisted (b) interface solitons. (c,d) Splitting of generated twisted interface solitons at the boundary between
a uniform and periodic media for (c) FF and (d) SH fields. Here $\beta=0$, $s=0.5$, and $K_1=K_2=4$.}
\label{fig4}
\end{figure}

Besides the odd solitons described above, we find that the tunable
waveguide with two dissimilar optical lattices also supports
higher-order interface lattice solitons, which can be viewed as
combinations of several odd solitons in engineered phases.
Fig.~\ref{fig3}(a) shows the case where the FF field features the
out-of-phase combination because the in-phase
combinations are unstable; we term the mode with an
out-of-phase combination as ``twisted interface lattice solitons''.
There exist cutoff values of propagation constant for twisted
interface solitons, which depend on the lattice period
[Fig.~\ref{fig3}(c)] and lattice depth [Fig.~\ref{fig3}(d)]. Note
that the profiles of twisted modes are asymmetric due to the
dissimilarity of the lattices on two sides [Fig.~\ref{fig3}(a)].
An important result is that such asymmetric twisted modes are
stable when the power exceeds a critical value [unstable region is
shown in dotted curves in Fig.~\ref{fig3}(b)], which is confirmed
by both, our linear stability analysis and the direct numerical
simulation of the soliton generation from the fundamental beam
in the framework of the model~(\ref{eq:model}). Generation and stable propagation
of asymmetric twisted surface solitons is also shown in
Fig.~\ref{fig4}(b).

\begin{figure}[h!]
\centerline{\includegraphics[width=7.0cm]{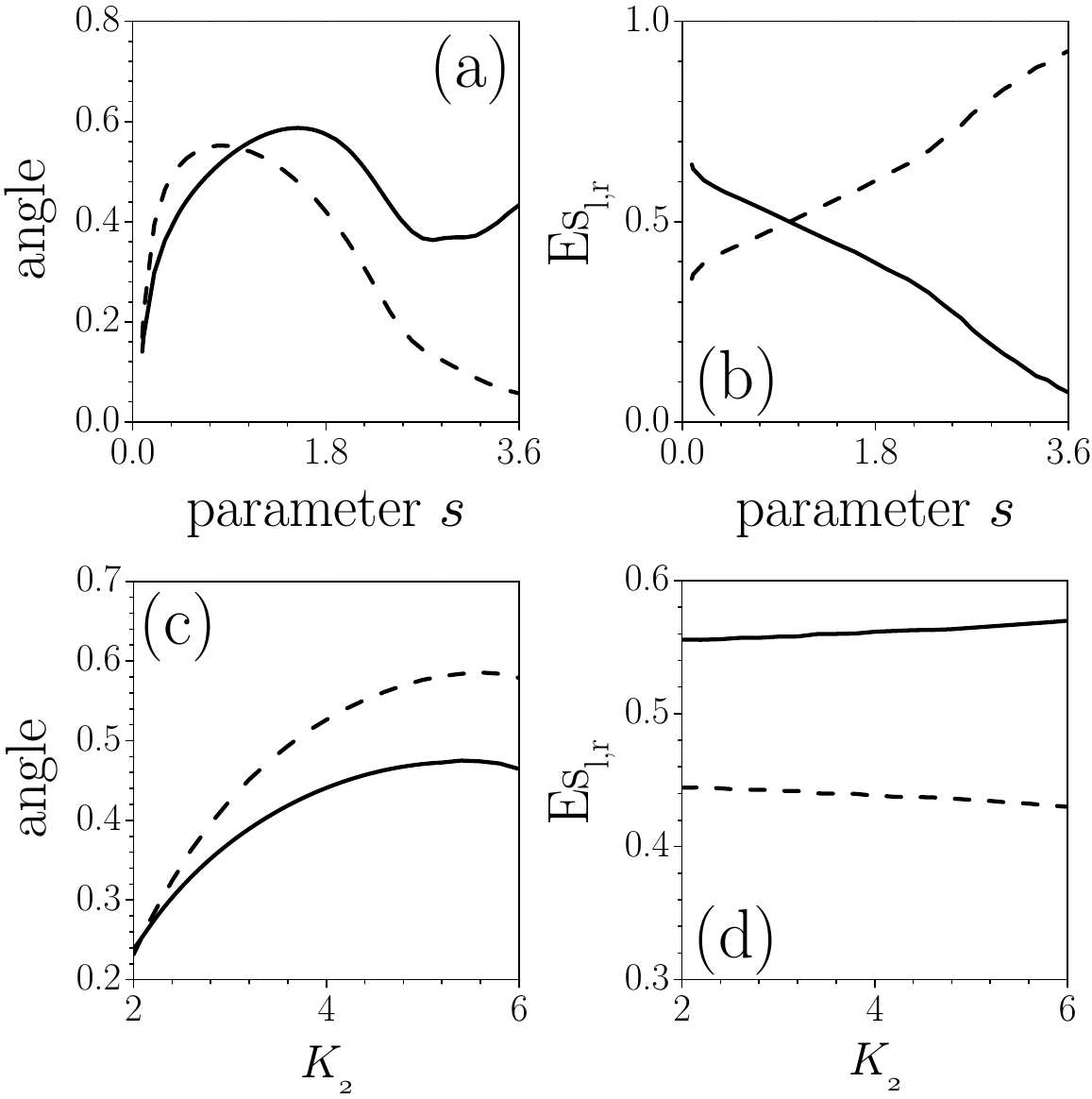}}
\caption{(a) Escape angles and (b) energy sharing of interface
twisted solitons as a function of parameter $s$ for $K_{2}=4$. (c)
Escape angles and (d) energy sharing of interface twisted solitons
as a function of parameter $K_{2}$ at $s=0.5$. Here Phase matching
$\beta=0$. In all cases solid and dashed curves correspond to the
left and right beams in the twisted soliton, respectively.}
\label{fig5}
\end{figure}

One of the key issues in this paper is to demonstrate the control
and manipulation of two-color interface solitons by tuning the
guiding parameters of the waveguide array. Removal of the
lattices causes each component of the twisted soliton to fly apart because the
interaction force between neighboring constituents is repulsive,
due to the out-of-phase combination that forms a twisted mode.
Figs.~\ref{fig4}(c,d) show a scenario in which a two-color interface
twisted soliton splits into two beams carrying different
energies when entering into a uniform medium from the quadratic
photonic lattices. Note that the power carried by each beam
is different due to the lattice-induced asymmetric profiles of
two constituents of the twisted solitons and as a result, the
escaping angles are different for the two beams. Thus, we can control
the output position of the two beams and the power sharing by tuning
the guiding parameters of the quadratic photonic lattices. The
whole manipulation and switching scenarios of twisted two-color
interface solitons in tunable lattices are summarized in
Fig.~\ref{fig5}, from which one can see clearly that the escape
angle and energies of two constituents of twisted solitons can be
controlled by tuning the relative lattice depth $s$ [Figs.~\ref{fig5}(a,b)]
and the modulation frequency $K_{2}$ of the lattices
[Figs.~\ref{fig5}(c,d)]. These results suggest novel opportunities
to control the flow of light in tunable photonic systems.

\section{Conclusions}

We have studied parametric localization of light at an interface separating two
%%%%%%%%%%%%%%%%%%%%%%%%%%%%%%
%%%%%%%%%%%%%%%%%%%%%%%%%%%%%%%%%%%%
different quadratically nonlinear photonic lattices and determined
the conditions for
the existence of interface localized modes. We have analyzed the impact of the phase mismatch on the properties
and stability of interface localized modes, as well as the
threshold power for their generation.
We have employed two different approaches: the coupled-mode theory and a continuous model with a periodic potential. Both models give basically the same qualitative results. Our results reveal that
asymmetric phase-twisted two-color surface solitons are stable in
a broad region of system parameters, but they can split into two solitons when entering a homogeneous medium. We have demonstrated that the
output position and energies of constituents of the two-color
interface solitons can be controlled by tuning the lattice and
waveguide parameters.

\section{Acknowledgements}

This work was supported by Fondecyt grant 1080374 and by the Australian Research Council.


\begin{thebibliography}{10}

\bibitem{Borstel} G. Borstel and H.J. Falge, in: {\em Electromagnetic Surface
Modes}, Ed. A.D. Boardman (Wiley, Chichester, 1982).
%, pp. 219-248.

\bibitem{Yeh_APL_78} P. Yeh, A. Yariv, and A.Y. Cho,  Appl. Phys. Lett. \textbf{32}, 104
(1978).

\bibitem{review} See, e.g., D. Mihalache, M. Bertolotti, and C. Sibilia, Prog. Opt.
\textbf{27}, 229 (1989), and references therein.

\bibitem{nw1} A.D. Boardman, P. Egan, F. Lederer, U. Langbein, and D.
Mihalache, in: \emph{Nonlinear Surface Electromagnetic Phenomena},
V.M. Agranovich, A.A. Maradudin, H.-E. Ponath, and G.I. Stegeman,
eds. (Elsevier Science Publishers B.V., New York, 1991), pp. 73-287.

\bibitem{OL_george} K.G. Makris, S. Suntsov, D.N. Christodoulides, G.I.
Stegeman, and A. Hach\'e, Opt. Lett. \textbf{30}, 2466 (2005).

\bibitem{PRL_george} S. Suntsov, K.G. Makris, D.N. Christodoulides, G.I.
Stegeman, A. Hach\'e, R. Morandotti, H. Yang, G. Salamo, and M.
Sorel, Phys. Rev. Lett. \textbf{96}, 063901 (2006).

\bibitem{PRL_kartashov} Ya.V. Kartashov, V.V. Vysloukh, and L. Torner,
Phys. Rev. Lett. \textbf{96}, 073901 (2006).

\bibitem{PRL_canberra} C.R. Rosberg, D.N. Neshev, W. Krolikowski, A.
Mitchell, R.A. Vicencio, M.I. Molina, and Yu.S. Kivshar, Phys. Rev.
Lett. \textbf{97}, 083901 (2006).

\bibitem{OL_inter} K.G. Makris, J. Hudock, D.N. Christodoulides, G.I. Stegeman, O.
Manela, M. Segev, Opt. Lett. {\bf 31}, 2774 (2006).

\bibitem{OE_poly} K. Motzek, A.A. Sukhorukov, and Yu.S. Kivshar, Opt. Exp. {\bf 14},
9873 (2006).

\bibitem{PLA_mario}  M.I. Molina and Yu.S. Kivshar, Phys. Lett. A
{\bf 362}, 280 (2007).

\bibitem{OE_stegeman01} S. Suntsov, K.G. Makris, D.N.
Christodoulides, G.I. Stegeman, R. Morandotti, M. Volatier, V.
Aimez, R. Ares, E. H. Yang, and G. Salamo, Opt. Express {\bf 16},
10480 (2008).

\bibitem{chi2_p1}  B.A. Malomed, P.G. Kevrekidis, D.J. Franzeskakis, H.E. Nistazakis,
and A.N. Yannacopoulos, Phys. Rev. E {\bf 65}, 056606 (2002).

\bibitem{chi2_p2} Ya.V. Kartashov, L. Torner, and V.A. Vysloukh, Opt. Lett. {\bf 29}, 1117
(2004);Ya.V. Kartashov, V.A. Vysloukh, and L. Torner, \emph{ibid}.
{\bf 29}, 1399 (2004).

\bibitem{chi2_p3} Z. Xu, Ya.V. Kartashov, L.-C. Crasovan, D. Mihalache, and L. Torner,
Phys. Rev. E {\bf 71}, 016616 (2005).

\bibitem{chi2_p4} H. Susano, P.G. Kevrekidis, R. Carretero-Gonzalez, B.A. Malomed,
and D.J. Frantzeskakis, Phys. Rev. Lett. {\bf 99}, 214103 (2007).

\bibitem{OE_stegeman} G.A. Siviloglou, K.G. Makris, R. Iwanow, R. Schiek, D.N. Christodoulides,
G.I. Stegeman, Y. Min, and W. Sohler, Opt. Express {\bf 14}, 5508 (2006).

\bibitem{chi2_our} Z. Xu and Yu.S. Kivshar, Opt. Lett. {\bf 33}, 2551 (2008).

\bibitem{OL_molina}
M. Molina, R. Vicencio, and Yu.~S. Kivshar, Opt. Lett. {\bf 31},
1693 (2006).

\end{thebibliography}
\end{document}